# Peculiarities in the pseudogap behavior in optimally doped YBa$_2$Cu$_3$O$_{7-\delta}$ single crystals under pressure up to 1 GPa

A. L. Solovjov$^{1,2}$, L. V. Omelchenko$^1$, R.V. Vovk$^3$, O. V. Dobrovolskiy$^4$, S. N. Kamchatnaya$^3$, D. M. Sergeyev$^{5*}$

$^1$*B. I. Verkin Institute for Low Temperature Physics and Engineering of National Academy of Science of Ukraine, 47 Nauki ave., 61103 Kharkov, Ukraine*
$^2$*International Laboratory of High Magnetic Fields and Low Temperatures, 95 Gajowicka Str., 53-421, Wroclaw, Poland*
$^3$*Physics Department, V. Karazin Kharkiv National University, Svobody Sq. 4, 61077 Kharkiv, Ukraine*
$^4$*Physikalisches Institut, Goethe University, Max-von-Laue-Str. 1, 60438 Frankfurt am Main, Germany*
$^5$*Zhubanov Aktobe Regional State University, 030000 Aktobe, Kazakhstan*

The influence of the hydrostatic pressure $P$ up to 0.95 GPa on the excess conductivity $\sigma'(T)$ and the pseudogap $\Delta^*(T)$ in optimally doped YBa$_2$Cu$_3$O$_{7-\delta}$ single crystals ($T_c \simeq 91.1$ K at ambient pressure) is investigated by electrical resistivity measurements. A pronounced enhancement of the pseudogap under pressure of $d\ln\Delta^*/dP \approx 0.32$, which is only a factor of 1.12 smaller than in slightly doped single crystals, is revealed for the first time. This implies a somewhat more moderate increase of the coupling strength in optimally doped cuprates with increasing pressure. Simultaneously, the ratio $2\Delta^*(T_c)/k_B T_c \approx 5$ at $P = 0$ GPa, which is typical for high-temperature superconductors with strong coupling, increases by 16% with increasing pressure. At the same time, the pressure effect on $T_c$ is minor: $dT_c/dP \approx +0.73$ KGPa$^{-1}$, whereas $d\ln\rho/dP \approx (-17 \pm 0.2)\%$ GPa$^{-1}$ is comparable with that in lightly doped YBCO single crystals. This suggests that the mechanisms of the pressure effect on $\rho(T)$ and $T_c$ are noticeably different. Independently of pressure, near $T_c$, $\sigma'(T)$ is well described by the Aslamazov-Larkin (3D-AL) and 2D Hikami-Larkin fluctuation theories, exhibiting a 3D–2D crossover with increasing temperature. However, the temperature interval $T_c < T < T_{01}$, in which $\sigma'(T)$ obeys the classical fluctuation theories, is exceptionally narrow ($\approx 1.16$ K). Nevertheless, a peculiarity at the temperature $T_{01}$, up to which the wave function phase stiffness in the superconductor is maintained, is clearly observed in the dependence $\Delta^*(T)$. Below $T_{01}$ a fast growth of $\Delta^*(T)$ is revealed for the first time. It can be associated with a sudden increase of the superfluid density, $n_s$, that is the density of fluctuating Cooper pairs (short-range phase correlations) forming in the sample when $T$ approaches $T_c$.

## 1. INTRODUCTION

In addition to a high critical temperature, $T_c$ [1], high-temperature superconductors (HTSCs) with active CuO$_2$ planes (cuprates) possess a series of other unconventional properties [2–6]. Among these are a low density of the charge carriers, $n_f$, which even in optimally doped samples is by an order of magnitude smaller than in conventional metals [6], strong electron correlations [7], quasi-two-dimensionality caused by the conductivity within the CuO$_2$ planes [8], and, in consequence of this, a strong anisotropy of the electronic properties [5–7, 9].

One of the most intriguing properties of cuprates is the pseudogap (PG) [5, 9–12] which is opening at some characteristic temperature $T^* \gg T_c$. By definition, PG is a state of matter characterized by a reduced (but nonzero) density of electronic states (DOS) at the Fermi level [13]. It should be stressed that the fundamental difference of the PG state from the superconducting (SC) one is that in the latter the SC gap is opening and DOS is equal to zero [9, 11]. It is expected that the correct understanding of the PG nature will allow one to shed light on the SC pairing mechanism in HTSCs which remains debatable so far. In particular, this is important for searching for novel superconductors with yet higher $T_c$'s. The number of theoretical models describing PG is exceptionally large [9, 11, 12, 14–16]. However, no ultimate elucidation of the PG nature is available so far. The majority of theorists assume that PG appears in consequence of fluctuations [17, 18]. However, so far there is no generally accepted viewpoint concerning the nature of these fluctuations (see [10, 12] and references therein). Some researchers believe that the appearance of PG is not directly associated with superconductivity [7, 10, 14–16, 19]. As possible sources for the appearance of PG, non-related to superconductivity, are considered such types of interaction as spin fluctuations [7, 20], charge-density waves [14], antiferromagnetic spin correlations [19], excitons [21], and polarons [22]. At the same time, the abundance of models points to that a generally accepted viewpoint in this regard is still absent.

The role of the electron-phonon interaction (EPI) in the formation of superconducting Cooper pairs at such high temperatures as well as of paired fermions (local pairs) above $T_c$ also remains unclear [23–25]. It is assumed that EPI should take place in cuprates [23], whereby EPI is probably enhanced by some additional interaction of, most likely, magnetic character [10, 15, 25]. However, recent calculations suggest [24] that the specific character of EPI in HTS's stipulates a strong correlational narrowing of the electron W-band. This leads to that the chemical potential $\mu \sim W \ll J$, where $J$ is the electron exchange interaction constant. The fulfillment of this condition is crucial for the formation of singlet electron pairs in HTSCs, coupled by a strong effective kinematic field. At the same time, the available experimental data do not suffice [25, 26] to examine the calculation results.

We adhere the viewpoint that PG in cuprates is caused by some specific fluctuations which lead to the formation of local pairs (LPs) at $T_c \leq T \leq T^*$, preceding the transition to the SC state [3, 9, 27–31]. According to theory [32–34], at high temperatures $T \leq T^*$ LPs appear in the form of strongly bound bosons (SBBs) obeying the Bose-Einstein condensation (BEC) theory. The SBB size is defined by the coherence length in the $ab$-plane, $\xi_{ab}(T) = \xi_{ab}(0)(T/T_c - 1)^{-1/2}$, whose spatial extent $\xi_{ab}(0) \sim \xi_{ab}(T^*)$ is exceptionally small. Thus, for YBCO films with a close-to-optimal doping level $\xi_{ab}(0) \sim 10\,\text{Å}$, typically [35, 36]. Accordingly, the coupling strength in such a pair, $\varepsilon_b \sim (\xi_{ab}^2)^{-1}$ is in turn exceptionally strong [32, 33]. Such a strong coupling, which can be caused by either of the above-mentioned interaction mechanisms and which is rigid against thermal fluctuations, ensures the possibility of the SBBs formation at such high temperatures. In consequence of this, SBBs are strongly coupled but local, i.e., non-interacting formations, since the pair size is much less than the distance between them. It should be noted that according to theory, SBBs can only appear in systems with a reduced $n_f$. This is, in particular, the case in cuprates with a doping level less than the optimal one. Importantly, PG is observed in HTSCs with just the mentioned doping level.

According to theory, LPs appear at some high temperature $T^*$ but they can only condensate at $T_c \ll T^*$ [32, 33]. This occurs in consequence of Gaussian fluctuations of the order parameter in two-dimensional (2D) systems to which HTSCs belong in a broad temperature range. Such fluctuations hinder the appearance of the phase coherence in the 2D state. As a result, $T_c$ in an ideal 2D metal turns out zero (Mermin-Wagner-Hohenberg theorem), and a finite value of $T_c$ can only be obtained when three-dimensional effects are taken into account [32, 37]. This is why, as our investigations of the fluctuation conductivity (FC) revealed [4, 9], HTSCs are always trimerizing when $T$ approaches $T_c$. At this conjuncture, FC is always described by the standard equation of the classical Aslamazov-Larkin theory [38] for 3D systems. However, by definition, non-interacting SBBs cannot condensate at all. This is why the theory [32, 38, 39] assumes that with decreasing temperature and simultaneously increasing $\xi_{ab}(T)$ SBBs should turn into fluctuational Cooper pairs (FCPs) obeying the BCS theory [40], i.e., it predicts the BEC-BCS crossover [32, 37]. The crossover temperature, $T_{pair}$, is clearly observed in a series of experiments [4, 41]. However, the crossover details are not completely clear so far.

Since the discovery of a strong dependence of $T_c$ on pressure in the compound La-Ba-Cu-O [42], pressure has been playing a rather noticeable role in investigations of HTSCs [43–46]. In contrast to conventional superconductors, the dependence $dT_c/dP$ in cuprates in the vast majority of cases is positive, whereas the derivative $d\ln\rho/dT$ is negative and rather large [44–46]. Here $\rho \equiv \rho_{ab}$ is the resistivity in the $ab$-plane of the sample, that is parallel to the $CuO_2$ planes. Upon application of pressure the volume of the elementary crystal cell is reduced. This has to contribute to the ordering of the system and should lead to a decrease of the number of structural defects and, hence, to a likely reduction of $\rho$. Nevertheless, the mechanisms of the pressure effect on $\rho$ still remain incomprehensible, since the nature of the transport properties of HTSCs is, strictly speaking, unclear. As is known, the major contribution to the conductivity in cuprates is provided by the $CuO_2$ planes, between which a relatively weak interplane interaction takes place. The application of pressure most likely leads to a rearrangement of the charge carriers resulting in an increase of the charge carriers density $n_f$ in the conducting $CuO_2$ planes and, hence, to a reduction of $\rho$. Apparently, an increase of $n_f$ under pressure should also lead to an increase of $T_c$ [6, 9], i.e., to a positive value of $dT_c/dP$, as observed in experiments.

Another mechanism, which can also raise $T_c$, refers to the possible increase of the pairing interaction $V_{eff}$ which has to depend on pressure. For underdoped cuprates the former mechanism is believed to dominate the pressure effect on $T_c$ (see Ref. [26] and references therein). It is well known that in YBCO the unique proximity between the $d$-Cu state and the $p$-O state is realized [47, 48]. As a result, the band structure of cuprate HTSCs is determined by the strongly correlated electron motion on the Cu(3d) orbital interacting with the O(2p) one. The applied hydrostatic pressure is very likely to affect this interaction. The effect of hydrostatic pressure on $\rho_{ab}$ in HTSCs was experimentally investigated in Ref. [44]. There is also a few works addressing the influence of pressure on FC in various cuprates [45, 46, 49, 50]. However, apart of a minor number of reports [51–53], there has been very few work devoted to the study of pressure on PG in HTSCs.

Here, we investigate the effect of the hydrostatic pressure $P \approx 1\,\text{GPa}$ ($1\,\text{GPa} = 10\,\text{kbar}$) on the temperature dependence of the resistivity $\rho_{ab}(T)$ in optimally doped $YBa_2Cu_3O_{7-\delta}$ single crystals with $T_c = 91.07\,\text{K}$ at $P = 0\,\text{GPa}$. We study the fluctuational contributions to the conductivity, focusing chiefly on the temperature dependence of the excess conductivity $\sigma'(T)$. From an analysis of the excess conductivity we obtain the PG temperature dependence $\Delta^*(T)$ at $P = 0\,\text{GPa}$ as well as at $0.25\,\text{GPa}$, $0.65\,\text{GPa}$, and $0.95\,\text{GPa}$. The analysis is performed in the framework of the local pair model (LP model) [4, 9], as detailed in the text. The reported comparison of our results with analogous ones for $Bi_2Sr_2CaCu_2O_{8-\delta}$ (BISCCO-2212)

[45], HgBa$_2$Ca$_2$Cu$_3$O$_8$ (Hg-2223) [49], HoBa$_2$Cu$_3$O$_{7-\delta}$ (HoBCO) [52] and slightly doped YBa$_2$Cu$_3$O$_{7-\delta}$ single crystals [26, 53] should contribute to a better understanding of the influence mechanisms of pressure on $T_c$, $\rho_{ab}(T)$, and $\Delta^*(T)$ in HTSCs.

## 2. EXPERIMENT

YBa$_2$Cu$_3$O$_{7-\delta}$ single crystals (YBCO) were grown by the solution-melt technique as described elsewhere [54–57]. For resistive measurements crystals of a rectangular shape with the typical sizes $3 \times 0.5 \times 0.03$ mm$^3$ were selected. The minimal crystal size corresponds to the $c$-axis. To obtain samples with a desired oxygen content, the crystals were annealed as detailed in Refs. [55–57]. The electrical resistance in the $ab$-plane was measured in the standard four-probe geometry [57]. The measurements were done in the temperature-sweep mode, with a rate of 0.1 K/min for measurements near $T_c$ and about 5 K/min for $T \gg T_c$. The hydrostatic pressure was created in a chamber of the cylinder-piston type [55, 56]. To exclude the influence of the oxygen redistribution, the measurements were done in two–seven days after the relaxation processes had been completed [52].

## 3. RESULTS AND DISCUSSION

### 3.1. Resistivity temperature dependence

The temperature dependences of the resistivity $\rho(T) = \rho_{ab}(T)$ of the YBa$_2$Cu$_3$O$_{7-\delta}$ single crystal with $T_c = 91.07$ K ($P = 0$ GPa) and the oxygen index $7-\delta \approx 6.94$ are shown in figure 1 at $P = 0$ GPa (curve 1) and $P = 0.95$ GPa (curve 2). The resistivity curves obtained for $P = 0.25$ GPa and $0.65$ GPa are in between these two curves and not shown for the sake of data readability. The shape of the $\rho(T)$ curves is typical for optimally doped YBCO films [58, 59] and single crystals [1, 60]. The transition temperature $T_c$ was determined by extrapolating the resistive transition to the value $\rho(T_c) = 0$ [61]. In the broad temperature range from 300 K to $T^* = (141 \pm 0.3)$ K ($P = 0$ GPa) and $T^* = (135.7 \pm 0.3)$ K ($P = 0.95$ GPa) the dependence $\rho(T)$ is linear with the slope $d\rho/dT = 0.63\,\mu\Omega\mathrm{cmK}^{-1}$ and $d\rho/dT = 0.54\,\mu\Omega\mathrm{cmK}^{-1}$ at $P = 0$ GPa and $P = 0.95$ GPa, respectively. It should be noted that as it follows from the theory and experiment [1, 6, 8, 9], in this case the values of the characteristic temperature $T^*$ are much smaller than $T^* \simeq 250$ K observed in slightly doped YBCO films [9, 58, 59] and single crystals [53, 60]. In addition to this, $T_c$'s are very high while the resistive transitions are exceptionally sharp, namely $\Delta T_c = T_c(0.9\rho'_N) - T_c(0.1\rho'_N) = 91.2$ K $-91.07$ K $= 0.13$ K ($P = 0$ GPa) and $\Delta T_c = 92$ K $-91.76$ K $= 0.24$ K ($P = 0.95$ GPa). Here, $\rho'_N$ is the resistivity at $T = T_{ons}$ corresponding to the onset of the SC transition

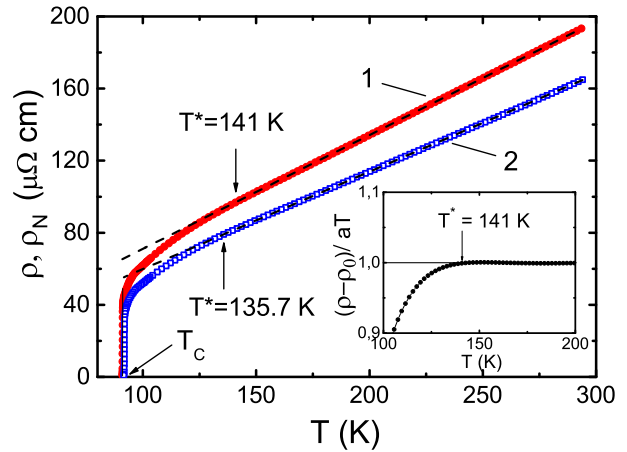

FIG. 1: Resistivity temperature dependences of the optimally doped YBa$_2$Cu$_3$O$_{7-\delta}$ ($7-\delta \approx 6.94$) single crystal at $P = 0$ GPa (curve 1) and $P = 0.95$ GPa (curve 2). Dashed lines designate $\rho_N(T)$ extrapolated to the low-temperature region. Inset: Determination of $T^*$.

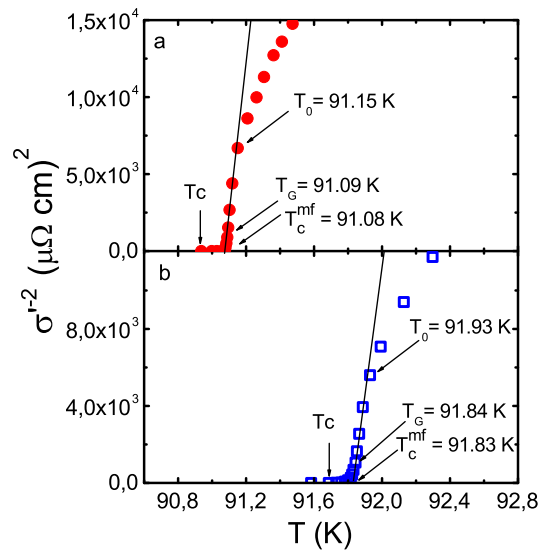

FIG. 2: Temperature dependences of $\sigma'^{-2}(T)$ of the YBa$_2$Cu$_3$O$_{6.94}$ single crystal at (a) $P = 0$ and (b) $P = 0.95$ GPa. Straight lines are guides for the eye.

[53]. Nevertheless, as in the case of slightly doped single crystals [53], pressure broadens the resistive transition by about a factor of two.

The applied pressure usually increases $T_c$ ($dT_c/dP \simeq +4$ KGPa$^{-1}$) and reduces $\rho(T)$ ($d\ln\rho/dP \simeq -12\%$ GPa$^{-1}$) of cuprates [44–46, 52]. In addition to this, it enhances the value of the SC gap, $\Delta(0)$, and the ratio $2\Delta/k_BT_c$ [25, 26], most likely due to a decrease of the phonon frequencies in cuprates under pressure [25]. In the case studied by us the *pressure* *effect on $T_c$ is very weak*: $dT_c/dP \simeq +0.73\,\text{KGPa}^{-1}$, see also figure 2. This is the first unexpected result revealed for optimally doped (OD) YBCO single crystals. For comparison, in slightly doped YBCO single crystals with $T_c = 49.2\,\text{K}$ we observed $dT_c/dP \simeq +5\,\text{KGPa}^{-1}$, in good agreement with the above-mentioned average value obtained for YBCO compounds with a relatively close-to-optimal doping level. However, as before, pressure strongly reduces the resistivity of the sample: $d\ln\rho/dP \approx -(17 \pm 0.2)\%\,\text{GPa}^{-1}$ that is even greater than the value obtained for Y123 [44] and of the same order of magnitude as reported for Bi2212 in the same reference. Hence, we can conclude that the mechanisms of the pressure effect on $T_c$ and $\rho$ are different. In optimally doped single crystals the charge carrier density in the $CuO_2$ planes, $n_f$, is maximal. Moreover, it is likely to saturate in YBCO compounds at the oxygen index $(7 - \delta \approx 6.94)$ we are dealing with in the studied sample. For this reason the pressure has practically no effect on $n_f$ as well as on $T_c$ related to it. Hence, we can conclude that the decrease of $\rho(P)$ is not related to the reduction of $n_f$ but it is likely to occur as a result of both, the reduction of the number of structural defects [44] and softening of the phonon spectrum [25] with increasing pressure. Also, we have to emphasize that for our slightly doped YBCO single crystal we observe $d\ln\rho/dP \approx -(19 \pm 0.2)\%\,\text{GPa}^{-1}$ which is by 12% larger than that found for the optimally doped one. The finding is confirmed by the PG results as will be shown below. This fact suggests a slightly weaker influence of the pressure on the YBCO phonon spectrum in the case of the OD samples.

### 3.2. Fluctuation conductivity

According to the theoretical model of nearly antiferromagnetic Fermi liquid (NAFL) [20], the linear dependence $\rho(T)$ at high temperatures corresponds to the normal state of the system. This state is characterized by the constancy of various possible interactions in HTSC and, hence, by the stability of the Fermi surface [10, 11, 20]. Below $T^*$, the measured $\rho(T)$ curve deviates from the linear dependence towards smaller values. This leads to the appearance of the excess conductivity

$$\sigma'(T) = \sigma(T) - \sigma'_N(T) = [1/\rho(T)] - [1/\rho_N(T)],$$
$$\sigma' = [\rho_N(T) - \rho(T)]/[\rho(T)\rho_N(T)], \quad (1)$$

where $\rho_N(T) = aT + b$ is the linear temperature dependence of the sample in the normal state [6, 8, 20]. By extrapolating $\rho_N(T)$ towards low temperatures and using equation (1), the temperature dependence $\sigma'(T)$ was determined. Using the LP model [4, 9], from the excess conductivity the data for FC and PG in the sample at pressures $P = 0\,\text{GPa}$, $0.25\,\text{GPa}$, $0.65\,\text{GPa}$, and $0.95\,\text{GPa}$ were deduced. Below we compare the results obtained for $P = 0\,\text{GPa}$ and $P = 0.95\,\text{GPa}$ applied during a week. The parameters obtained from the analysis for all samples studied are summarized in Table I. The first step in the LP model analysis requires to deduce the dependence $\ln\sigma'(\ln\varepsilon)$ shown in figure 3, which determines the range of superconducting fluctuations in the vicinity of $T_c$ [9, 38, 39]. Here, $\varepsilon = (T/T_c^{mf} - 1)$ is the reduced temperature and $T_c^{mf}$ is the critical temperature in the mean-field approximation [32, 62, 63], which separates the range of SC fluctuations from the range of critical fluctuations near $T_c$ (where $\Delta < k_B T$), not accounted for within the Ginzburg-Landau theory [64, 65].

For the determination of $T_c^{mf}$ one uses the experimental fact that in all HTSCs near $T_c$, $\sigma'(T)$ is always extrapolated by the 3D Aslamazov-Larkin (AL) equation [see equation (2) below] [9, 35, 36, 58, 59, 61]. In this equation $\sigma'(T)$ diverges as $\varepsilon^{-1/2}$ when the temperature is close to $T_c^{mf}$. Accordingly, $\sigma'^{-2} \sim \varepsilon \sim (T - T_c^{mf})$ [66], that allows one to determine $T_c^{mf}$ and, hence, $\varepsilon$. In figure 2 we present the dependence $\sigma'^{-2}(T)$ of the investigated sample at $P = 0\,\text{GPa}$ (panel (a), full circles) and $P = 0.95\,\text{GPa}$ (panel (b), open squares). The intersection of the linear dependence $\sigma'^{-2}(T)$ with the temperature axis yields $T_c^{mf} = 91.08\,\text{K} > T_c = 91.07\,\text{K}$ ($P = 0\,\text{GPa}$) and $T_c^{mf} = 91.83\,\text{K} > T_c = 91.76\,\text{K}$ ($P = 0.95\,\text{GPa}$), respectively. In addition to this, in figure 2 we mark the 3D-2D (AL-MT) crossover temperature $T_0$ and the Ginzburg temperature $T_G$, up to which the experimental data obey the AL theory when $T$ approaches $T_c^{mf}$ [67, 68]. In figure 3 this temperature corresponds to the marked value of $\ln(\varepsilon_G)$.

The dependence $\ln\sigma'(\ln\varepsilon)$ is displayed in figure 3 for $P = 0\,\text{GPa}$ (panel (a), full circles) and $P = 0.95\,\text{GPa}$ (panel (b), open squares). Both curves are noticeably shifted to the left as compared to the slightly doped YBCO single crystals ($T_c = 49.2\,\text{K}$ at $P = 0\,\text{GPa}$) [53]. This points to the smallness of the coherence length $\xi(T) = \xi(0)(T/T_c^{mf} - 1)^{-1/2}$ [63] in the sample. Nevertheless, up to $T_0 \approx 91.15\,\text{K}$ ($\ln\varepsilon_0 \approx -7.11$, $P = 0\,\text{GPa}$) the experimental data are fitted well to the AL fluctuation contribution for 3D systems [38]

$$\sigma'_{AL3D} = C_{3D}\frac{e^2}{32\,\hbar\,\xi_c(0)}\varepsilon^{-1/2}. \quad (2)$$

which is presented by the dashed straight line with the slope $\lambda = -1/2$ in figure 3, and above $T_0$ up to $T_{01} \approx 92.54\,\text{K}$ ($\ln\varepsilon_{01} \approx -4.2$) by the Maki-Thomson (MT) contribution of the Hikami-Larkin (HL) theory [39]

$$\sigma'_{MT} = \frac{e^2}{8\,d\,\hbar}\frac{1}{1 - \alpha/\delta}\,ln\left((\delta/\alpha)\frac{1 + \alpha + \sqrt{1 + 2\,\alpha}}{1 + \delta + \sqrt{1 + 2\,\delta}}\right)\varepsilon^{-1}, \quad (3)$$

see the solid curve in figure 3(a).

Analogous results have been obtained for all other pressures, including $P = 0.95\,\text{GPa}$, refer to figure 3(b), for which $T_0 \approx 91.93\,\text{K}$ ($\ln\varepsilon_0 \approx -6.72$) and $T_{01} \approx 93.18\,\text{K}$

Parameters of the optimally doped YBa$_2$Cu$_3$O$_{6.94}$ single crystal.

TABLE I:

| P | $\rho(280K)$, | $T_c$, | $T_c^{mf}$, | $T_{01}$, | $T_G$, | $\Delta T_{fl}$, | $d_1$, | $\xi_c(0)$, | $T^*$, | $T_{pair}$, | $\Delta^*(T_G)$, | $\dfrac{2\Delta^*}{k_B T_c}$ | $\varepsilon_{c0}^*$, | $\alpha_0$ |
|---|---|---|---|---|---|---|---|---|---|---|---|---|---|---|
| GPa | $\mu\Omega$cm | K | K | K | K | K | Å | Å | K | K | K | | | |
| 0    | 184.6 | 91.07 | 91.08 | 92.54 | 91.09 | 1.45 | 2.72 | 0.334 | 141   | 129 | 228 | 5   | 0.154 | 0.53 |
| 0.25 | 175.9 | 91.12 | 91.3  | 92.6  | 91.3  | 1.3  | 2.73 | 0.335 | 135   | 127 | 239 | 5.2 | 0.151 | 0.53 |
| 0.65 | 168.2 | 91.51 | 91.58 | 92.9  | 91.6  | 1.3  | 3.34 | 0.47  | 140   | 126 | 230 | 5.4 | 0.143 | 0.54 |
| 0.95 | 157.3 | 91.76 | 91.83 | 93.18 | 91.84 | 1.34 | 3.3  | 0.405 | 135.7 | 122 | 273 | 5.8 | 0.147 | 0.53 |

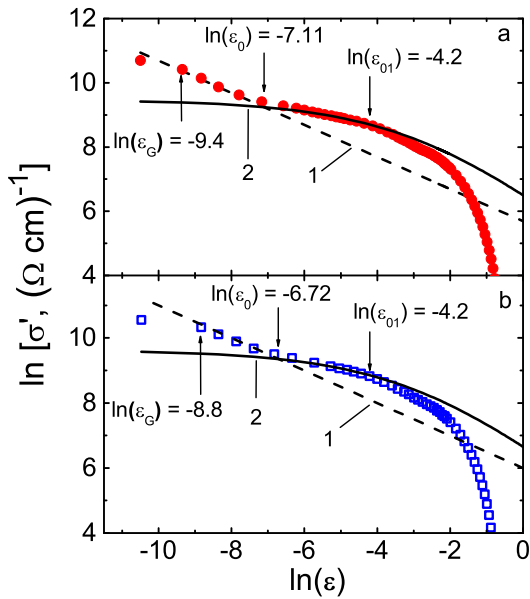

FIG. 3: $\ln\sigma'$ vs $\ln\varepsilon$ of the YBa$_2$Cu$_3$O$_{6.94}$ single crystal at $P = 0$ GPa (panel a, dots) and $P = 0.95$ GPa (panel b, open squares) compared with the fluctuation theories: 3D AL (dashed lines 1) and 2D MT (solid curves 2). $\ln\varepsilon_{01}$ corresponds to $T_{01}$ which determines the range of SC fluctuations, $\ln\varepsilon_0$ corresponds to the crossover temperature $T_0$, and $\ln\varepsilon_G$ designates the Ginzburg temperature $T_G$.

($\ln\varepsilon_{01} \approx -4.2$). As usually, in equations (2) and (3) $\alpha = 2[\xi_c(0)/d]^2\varepsilon^{-1}$ is the pairing parameter,

$$\delta = \beta\frac{16}{\pi\,\hbar}\left[\frac{\xi_c(0)}{d}\right]^2 k_B T\,\tau_\phi \qquad (4)$$

is the depairing parameter, and $\tau_\phi$ is the fluctuating Cooper pairs relaxation time defined as $\tau_\phi\beta T = (\pi\hbar)/(8k_B\varepsilon_{01}) = A/\varepsilon_{01}$. Here, $\beta = 1.203(l/\xi_{ab})$, where $l$ is the mean free path and $\xi_{ab}$ is the coherence length in the $ab$-plane in the clean-limit approximation [9].

At $T < T_0$, that is near $T_c$, the coherence length along the $c$-axis $\xi_c(T) > d$, where $d = 11.67$ Å is the lattice parameter along the $c$-axis. Accordingly, FCPs can interact in the entire volume of the superconductor, thereby forming the 3D state. That is, HTSCs are always trimerized near the SC transition, in accordance with theory [32–34, 37]. Above $T_0$, $d > \xi_c(T) > d_{01}$, where $d_{01}$ is the distance between the conducting CuO$_2$ planes. This means that the 3D state is lost, but the Josephson interaction still couples the neighboring CuO$_2$ planes [8, 39, 69]. Hence, this is a quasi-two-dimensional state of the system, which is described by the 2D-MT fluctuation contribution of the HL theory by equation (3). Consequently, $T_0$ is the MT-AL transition temperature and, simultaneously, the 2D-3D crossover temperature. Evidently, $\xi_c(T_0) = d$, that allows one to deduce

$$\xi_c(0) = d\sqrt{\varepsilon_0}. \qquad (5)$$

At $P = 0$ GPa equation (5) yields $\xi_c(0) = (3.34 \pm 0.02) \times 10^{-1}$ Å that is very small. One can see in figure 3(b) that in this case pressure affects the value of $\sigma'(T)$ only slightly. Nevertheless, at $P = 0.95$ GPa $\xi_c(0) = (4.05 \pm 0.02) \times 10^{-1}$ Å, i.e., the coherence length slightly grows under pressure, as in other cuprates [49]. Usually, in cuprates the coherence length in the $ab$-plane, which determines the Cooper pair size, $\xi_{ab}(0) \sim 15\xi_c(0)$. Hence, in our optimally doped single crystal with $T_c \sim 91$ K $\xi_{ab}(0) \approx 5$ Å at $P = 0$ GPa is expected and this value is in good agreement with our finding. For comparison, $\xi_{ab}(0)$ is about 13 Å in slightly under-doped YBCO with $T_c = 87.4$ K [9, 35, 36, 58, 59]. In this way, we have got the second nontrivial result, namely a *very small pair size at high temperatures* where LPs should exist in the form of SBBs. At the same time, as mentioned before, the smaller $\xi_{ab}(0)$, the larger the bound energy in the pair $\varepsilon_b \sim (\xi_{ab}^2)^{-1}$ [32, 33, 62]. This estimate appears reasonable taking into account the very high $T_c$ of the investigated crystal.

Another characteristic temperature in figure 3 is $T_{01}$. It determines the range of the SC fluctuations above $T_c$. Having determined $\ln\varepsilon_{01}$ from the data one gets $T_{01} \approx 92.54$ K ($\ln\varepsilon_{01} \approx -4.2$ at $P = 0$ GPa) and $T_{01} \approx 93.2$ K ($\ln\varepsilon_{01} \approx -4.2$ at $P = 0.95$ GPa). As mentioned above, at $T_0 < T < T_{01}$, $\xi_c(T) < d$, but simultaneously $\xi_c(T) > d_{01}$, that is the system is in the quasi-2D state and it is described by equation (3) [39, 69]. Accordingly, above $T_{01}$, where $\xi_c(T) < d_{01}$, the pairs are located within the CuO$_2$ planes and do not interact. This is why above $T_{01}$ the fluctuation theories do not describe the experiment, as is clearly seen in figure 3. In this way, it follows that $\xi_c(T_{01}) = d_{01}$ at $T = T_{01}$. It is evident that $\xi_c(0) = const$ at a given pressure, so that the condition $\xi_c(0) = d_{01}\sqrt{\varepsilon_{01}}$ should be fulfilled. Since $\xi_c(0) = 3.34 \times 10^{-1}$ Å is determined by the temperature of the dimensional crossover $T_0$, see equation (5), this allows one to estimate $d_{01} = \xi_c(0)(\sqrt{\varepsilon_{01}})^{-1}$. This yields $d_{01} \approx 2.8$ Å at $P = 0$ GPa and $d_{01} \approx 3.3$ Å at $P = 0.95$ GPa that is close to the values of $d_{01}$ determined from structural studies of YBCO [1, 70]. In this way, despite the very small values of $\xi_c(0)$, the analysis of the excess conductivity in the framework of the LP model allows one to get reasonable values of $d_{01}$. However, in contradistinction to lightly doped YBCO single crystals [53], $d_{01}$ grows slightly with pressure. This may be caused by the error in the determination of $\ln \varepsilon_{01}$ in figure 3 or by the peculiarities in the behavior of optimally doped single crystals under pressure.

On the other hand, according to theory [27], it is this temperature $T_{01}$ up to which the order parameter phase stiffness is maintained in HTSCs, as confirmed in experiments [71]. This means that in the temperature range from $T_c$ to $T_{01}$ Cooper pairs chiefly behave as superconducting pairs. This leads to a behavior of the cuprates, which is unconventional from the viewpoint of "classical" superconductivity. As it was shown in a series of works [3, 28, 70], the SC gap in HTSCs does not vanish at $T_c$ and the range of SC fluctuations is maintained up to $\sim 120$ K in YBCO ($\approx 30$ K above $T_c$) and up to $\sim 150$ K in Bi2223 ($\approx 40$ K above $T_c$). For instance, in the slightly doped YBCO single crystal investigated by us $T_c = 49.2$ K and $T_{01} = 85.2$ K [53]. This means that the interval, in which $\sigma'(T)$ is described by the fluctuation theories, that is where the phase stiffness of the order parameter is maintained, $\Delta T_{fl} \approx 36$ K, is in good agreement with the aforementioned results.

In the considered optimally doped single crystal, all temperature intervals, in which $\sigma'(T)$ can be described by the fluctuation theories, are exceptionally narrow. In particular, $\Delta T_{fl} = T_{01} - T_G = 92.54$ K $-91.09$ K $= 1.45$ K. This is in good agreement with the theory, namely the higher $T_c$, the narrower the range of SC fluctuations [10, 11, 27]. In addition, the analogous result follows from the analysis of the temperature dependence of the pseudogap $\Delta^*(T)$ (Figs. 6 and 7), where we have succeeded to distinctly observe a peculiarity corresponding to $T_{01}$ at all values of the applied pressure, as detailed next.

### 3.3. Analysis of the pseudogap temperature dependence

Since the excess conductivity $\sigma'(T)$, equation (1), is believed to appear due to PG formation [4, 9, 11], it has to contain information about the PG. To deduce this information, one evidently needs an equation which describes $\sigma'(T)$ over the whole temperature range $T_c < T < T^*$ and contains the PG parameter $\Delta^*$ in the explicit form. The equation for $\sigma'(T)$ accounting for the LP formation in HTSCs at $T \leq T^*$ reads [4, 9]

$$\sigma'(\varepsilon) = \frac{e^2 A_4 \left(1 - \dfrac{T}{T^*}\right)\left(\exp\left(-\dfrac{\Delta^*}{T}\right)\right)}{(16\,\hbar\,\xi_c(0)\,\sqrt{2\,\varepsilon_{c0}^*}\,\sinh(2\,\varepsilon\,/\,\varepsilon_{c0}^*)}. \quad (6)$$

Here $(1 - T/T^*)$ stands for the number of pairs formed at $T < T^*$, while $(\exp(-\Delta^*/T))$ takes into account the number of pairs broken by fluctuations when $T$ approaches $T_c$. Solving equation (6) with respect to the parameter $\Delta^*$, which describes the pseudogap, one gets

$$\Delta^*(T) = T \ln \frac{e^2 A_4 \left(1 - \dfrac{T}{T^*}\right)}{\sigma'(T)\,16\,\hbar\,\xi_c(0)\,\sqrt{2\,\varepsilon_{c0}^*}\,\sinh(2\,\varepsilon\,/\,\varepsilon_{c0}^*)}. \quad (7)$$

where $\sigma'(T)$ is the excess conductivity measured in experiment. In both equations $T$ is the given temperature and $T^*$ is the pseudogap opening temperature determined from the resistivity measurements (figure 1). $\xi_c(0)$ is the coherence length along the c-axis, which is determined by the 2D-3D crossover temperature $T_0$ from the temperature dependence of the fluctuation conductivity (figure 3). In this way, for the determination of $\Delta^*(T)$ one has to define from experiment the theoretical parameter $\varepsilon_{c0}^*$ and the scaling coefficient $A_4$. This is easily accomplished in the framework of the LP model [4, 9], as detailed next.

It turned out [72] that in cuprates, in the temperature range $\varepsilon_{c01} < \varepsilon < \varepsilon_{c02}$ ($T_{c01} < T < T_{c02}$) the value $\sigma'^{-1} \propto \exp(\varepsilon)$, that is the dependence $\ln(\sigma'^{-1})$ on $\varepsilon$ is linear (inset to figure 4). The parameter $\varepsilon_{c0}^* = 1/\alpha^* = 0.154$ [4, 9, 72] is determined from the slope of the straight line $\alpha^* = 6.5$. To calculate $A_4$, one has to plot the $\ln \sigma'(\ln \varepsilon)$ in the whole range from $T^*$ to $T_c^{mf}$, as shown in figure 4 (symbols) for $P = 0$ GPa. Then, using equation (6) with the already found parameters one calculates the dependences $\ln \sigma'(\ln \varepsilon)$. Finally, by varying $A_4$ the calculated curve has to be fitted to the experimental data $\ln \sigma'(\ln \varepsilon)$ in the range of 3D AL fluctuations near $T_c$ [4, 9] (figure 4, dashed line), where it is assumed that $\Delta^*(T_c) = \Delta(0)$ [28, 73]. For this one also has to know the value of $\Delta^*(T_c)$ in equation (6).

In order to find $\Delta^*(T)$ one has to plot the experimental $\ln \sigma'$ versus $1/T$ [9, 74] [figure 5(a) and 5(b), symbols] and to fit them to $\ln \sigma'(1/T)$ calculated by equation (6), figure 5, solid curves. With this procedure the shape of the curves determined by equation (6) turns out very sensitive to the $\Delta^*(T_c)$ value. As it follows from the figure, the best fit ensues at $D^* = 2\Delta^*(T_c)/k_B T_c = 5$ ($P = 0$ GPa) and $D^* = 5.8$ ($P = 0.95$ GPa) that points to the strong coupling limit. This result appears reasonable, taking into account the fact that the sample is optimally doped with a very high $T_c \sim 91.1$ K. One can also see that the pressure enhances the value of $D^*$ by 16%, in agreement with Refs. [25, 26] reported the enhancement of the SC gap $\Delta$ and the ratio $2\Delta(0)/k_B T_c$ under pressure. One more distinctive feature of the optimally doped samples with respect to the slightly doped single crystals is that equation (6) poorly describes the experimental values of $\sigma'(T)$ above $T_0$ (figure 4). However, should one substitute in equation (6) the $\Delta^*(T)$ data reported in figure 6 instead of the $\Delta^*(T_c)$ values, equation (6) will describe the experiment very well.

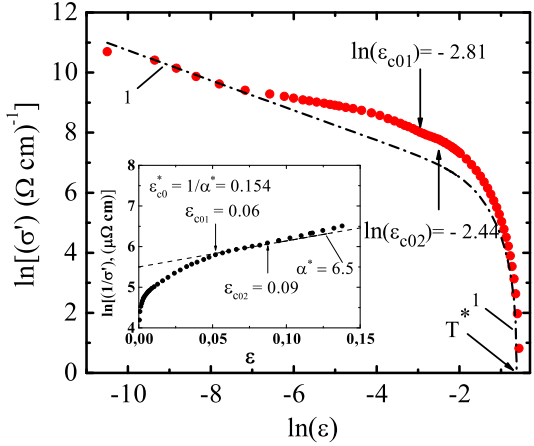

FIG. 4: Fluctuation conductivity $\ln \sigma'$ as a function of $\ln \varepsilon$ (symbols) of the $YBa_2Cu_3O_{6.94}$ single crystal at $P = 0\,GPa$ plotted in the whole temperature range from $T^*$ down to $T_c^{mf}$. Dashed curve 1 — equation (6) with the parameters detailed in the text. Inset: $\ln(1/\sigma')$ versus $\varepsilon$ which is linear between $\varepsilon_{c01} = 0.06$ and $\varepsilon_{c02} = 0.09$. The corresponding $\ln \varepsilon_{c01} \simeq -2.81$ and $\ln \varepsilon_{c02} \simeq -2.44$ are marked by the arrows in the main panel. The slope of the linear part $\alpha^* = 0.65$ determines the parameter $\varepsilon_{c0}^* = 1/\alpha^* = 0.154$.

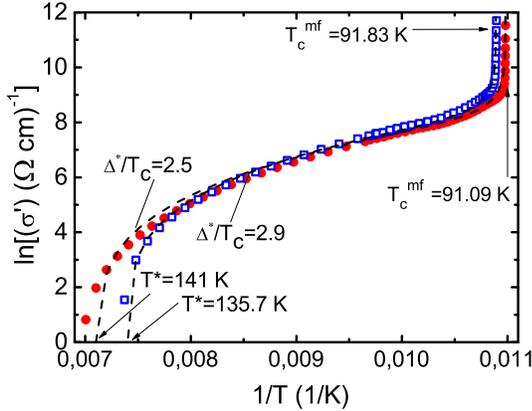

FIG. 5: Fluctuation conductivity $\sigma'$ as a function of the $1/T$ of $YBa_2Cu_3O_{6.94}$ single crystal at $P = 0\,GPa$ (full circles) and $P = 0.95\,GPa$ (open squares). Dashed curves — calculation by equation (6) with the parameters detailed in the text.

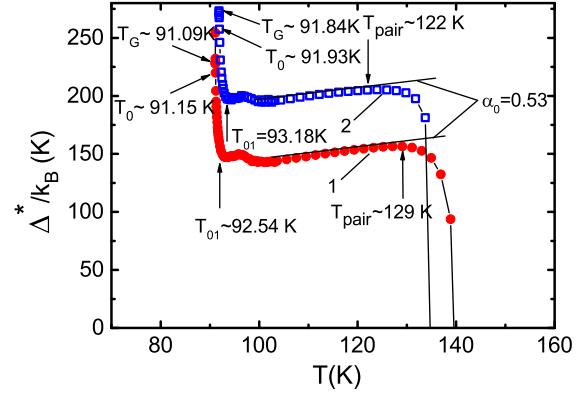

FIG. 6: Temperature dependences of the pseudogap $\Delta^*(T)$ in the $YBa_2Cu_3O_{6.94}$ single crystal at $P = 0\,GPa$ (curve 1) and $P = 0.95\,GPa$ (curve 2) calculated within the local pair model by equation (7) with the parameters detailed in the text. Solid curves are guides for the eye.

When all the parameters required are found, one can plot the dependences $\Delta^*(T)$ for all values of $P$. Curve 1 in figure 6 displays the dependence $\Delta^*(T)$ obtained within the LP model by equation (7) at $P = 0\,GPa$ with the following parameters deduced from the experiment: $T^* = 141\,K$, $T_c^{mf} = 91.08\,K$, $\xi_c(0) = 0.334\,Å$, $\varepsilon_{c0}^* = 0.154$, and $A_4 = 4.7$. The respective dependence obtained at $P = 0.95\,GPa$ is shown in figure 6 by open squares (curve 2). It is plotted with the parameters $T^* = 135.7\,K$, $T_c^{mf} = 91.83\,K$, $\xi_c(0) = 0.405\,Å$, $\varepsilon_{c0}^* = 0.147$, and $A_4 = 12$. The corresponding dependences for $P = 0.25\,GPa$ and $0.65\,GPa$ are between these two curves but not shown to for the sake of data readability. The respective parameters obtained within the LP model analysis at $P = 0.25\,GPa$ and $0.65\,GPa$ are summarized in Table I. It is worth to note that, as well as in slightly doped YBCO single crystals [53], there is a peculiarity in the sample pressure behavior at $P \approx 6.5\,GPa$. This peculiarity is seen in the pressure dependence of $d_{01}$ and $\xi_c(0)$, as well as of $\varepsilon_{c0}^*$, $D^*$, $\Delta^*(T_G)$ and $T_{pair}$, whereas $\rho(T)$ and $T_c$ smoothly vary with pressure.

Figure 6 shows that pressure noticeably increases $\Delta^*$ at a rate $d\ln \Delta^*/dP \approx 0.32$. Simultaneously, the ratio $D^* = 2\Delta^*(T_c)/k_B T_c$ also increases by 16%, whereas the shape of the $\Delta^*(T)$ curve is modified only slightly. Indeed, independently of pressure, the pseudogap $\Delta^*(T)$ sharply increases in the range $T^* > T > T_{pair}$ demonstrating maximum at $T_{pair} \simeq 129\,K$ ($P = 0\,GPa$), which is very close to $T_{pair} \approx 130$ K being typical for high-quality YBCO thin films with different oxygen concentrations [9, 75]. $T_{pair}$ corresponds to the temperature, at which LPs transform from SBBs into FCPs [4, 9], as mentioned above. Below $T_{pair}$ the dependence $\Delta^*(T)$ becomes linear with a positive slope $\alpha_1 \approx 0.53$, which appears to be pressure-independent. However, as well as in slightly doped YBCO single crystals [53], $T_{pair}$ somewhat decreases under pressure down to $\approx 122$ K at $P = 0.95$ GPa, see figure 6. Regardless of the $P$ value, linearity is maintained down to $\approx 100$ K. A feeble maximum at $T \approx 96$ K is likely caused by a feature of the studied sample, namely, by the peculiarity in $\rho(T)$ at $T \approx 96$ K, which is also seen as a minimum in the dependence $\ln \sigma'(\ln \varepsilon)$ at $\ln \varepsilon \approx -2.6$ (figure 3).

Independently of pressure, below $T_{01}$ down to $T_G \geq T_c^{mf}$ a *sudden growth of the pseudogap* $\Delta^*(T)$ takes place, which is distinctly observed for the first time. This is the third non-trivial result reported in this work. Such a behavior is most likely stipulated by a transition of the sample into the range of SC fluctuations, refer to figure 3. The specific character of the HTSCs behavior, as mentioned above, consists in the that the wave function phase stiffness of the superconductor is maintained just up to $T_{01}$ [27, 71]. This means that the superfluid density, $n_s$, maintains a nonzero value above $T_c$ up to $T_{01}$, i. e., the fluctuating pairs below $T_{01}$ behave like conventional SC pairs [27, 71, 76]. However, in contrast to slightly doped YBCO single crystals [52], where the interval of SC fluctuations $\Delta T_{fl} = T_{01} - T_G \approx 36$ K, that is in line with the results of Refs. [71, 76], in the case of OD single crystals $\Delta T_{fl} = T_{01} - T_G = (92.54 - 91.09)$ K = $1.45$ K ($P = 0$ GPa) and $\Delta T_{fl} = 1.34$ K ($P = 0.95$ GPa), that is exceptionally narrow. Nevertheless, this result is in agreement with the phase diagram of HTSCs (see Refs. [1, 10] and references therein), namely, the higher the charge carrier density $n_f$ in the sample, the higher $T_c$ and the lower $T^*$, and, as one can see now, the narrower the range of the SC fluctuations above $T_c$. In any case, the character of the $\Delta^*(T)$ dependence markedly changes at $T = T_{01}$ most likely due to the dramatic change of the LP interaction in the sample at this temperature.

The growth of $\Delta^*(T)$ at $T < T_{01}$ is presented in figure 7 in detail ($P = 0$ GPa). One sees that similarly to slightly doped YBCO single crystals [53], $\Delta^*(T)$ rapidly increases at $T < T_{01}$, has a maximum near $T_0$ and a minimum at $T_G$, below which there is a transition to the range of critical fluctuations [9], and the fluctuation theories fail. In this way, one can conclude that the transition to the SC state in both, slightly doped (SD) and optimally doped (OD) YBCO single crystals, occurs in identical way. The only exception is the disappearance of the maximum between $T_0$ and $T_G$ in the OD single crystals at $\sim 1$ GPa. However, one should emphasize two additional essentially distinctive features. In SD single crystals the range of SC fluctuations is very broad ($\Delta T_{fl} \approx 36$ K) whereas the increase $\Delta^*(T) = \Delta^*(T_G) - \Delta^*(T_{01}) \approx 5$ K ($P = 0$ GPa) is rather small [52]. It is worth to emphasize that a similar increase of $\Delta^*$ near $T_c$ is observed for many different HTSCs including HoBCO single crystals [52] and FeAs-based superconductors [77]. This fact suggests that this increase of PG below $T_{01}$ is most likely a typical feature of the HTSCs behavior just before the SC transition. The specific character of the OD single crystals consists in the fact that the increase $\Delta^*(T) = \Delta^*(T_G) - \Delta^*(T_{01}) \approx 80$ K at $P = 0$ GPa is very large but occurs in the exceptionally narrow temperature interval $\Delta T_{fl} \approx 1.5$ K (Figs. 6 and 7).

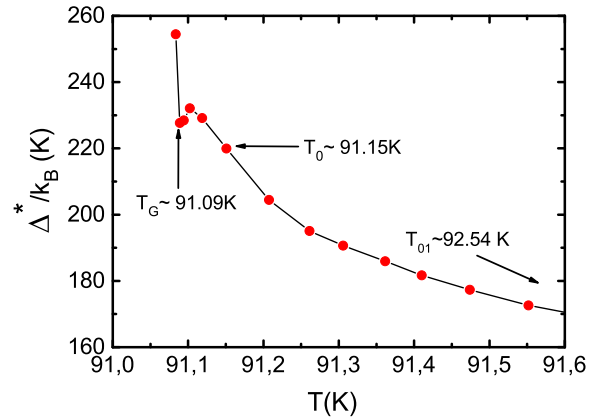

FIG. 7: Temperature dependence of the pseudogap $\Delta^*(T)$ in the optimally doped YBa$_2$Cu$_3$O$_{6.94}$ single crystal at $P = 0$ GPa (symbols) at $T < T_{01}$. Solid curve is guide for the eye.

Such s pronounced increase of $\Delta^*(T)$ at $T < T_{01}$ is likely stipulated by a rapid increase of the number of coherent FCPs (short-range phase correlations) below $T_{01}$. This conclusion can be confirmed by the results of frequency-dependent complex conductivity $\sigma(\omega)$ measurements on slightly doped Bi$_2$Sr$_2$CaCu$_2$O$_{8+\delta}$ (BSCCO) films with $T_c \approx 74$ K [71]. Below $T_c$, the conductivity measures the phase-stiffness energy $k_B T_\theta$ directly, viz.,

$$\sigma(\omega) = i\sigma_Q (k_B T_\theta / h\omega), \tag{8}$$

where $\sigma_Q \equiv e^2/hd$ is the quantum conductivity of a stack of planar conductors with the interlayer spacing $d$. It was shown [71] that $\sigma(\omega) \propto T_\theta$ persists up to $T_{01} \approx 100$ K, more than 25 K above $T_c$, suggesting the expected presence of the short-range phase correlations in HTSCs well above $T_c$. In optimally doped YBCO samples the density of the charge carries $n_f$ is at least a factor of 3 larger than in SD single crystals with $T_c \approx 49$ K [6, 8, 58, 59]. Accordingly, the density of superconducting carriers, $n_s$, and, consequently, the density of forming FCPs (short-range phase correlations) above $T_c$ should also be noticeably larger, that can explain the observed marked increase of $\Delta^*$ at $T < T_{01}$ in this case. Besides, the very short $\xi(T)$ obtained for the OD samples is likely to promote the forming of the FCPs correlation. It should also be noted that the absolute value of $\Delta^*(T_G) \approx 230$ K (Figs. 6 and 7) is a factor of 2 larger than in the SD single crystals.

Finally, it has to be emphasized that increase of $\Delta^*$ observed in OD YBCO under pressure is a factor of 1.12

less than that found for SD single crystals [53]. Note that just the same result demonstrates the resistivity, namely, the decrease of $\rho$ is also a factor of 1.12 less than in the SD samples. Thus, one may conclude that both effects are based on the same physical principle which can be accounted for in terms of the electron-phonon interaction. As convincingly shown in Ref. [25], a noticeable increase of both, the SC gap $\Delta(0)$ and the ratio $2\Delta(0)/k_B T_c$ with pressure, observed in slightly doped polycrystals Bi2223, is accompanied by a marked decrease of the phonon spectrum frequencies in the superconductor. In this way, it is likely that the observed phonon spectrum softening is the primary cause for the increase of both, the SC gap [25, 26] and the pseudogap $\Delta^*$ ([53] and this paper) under pressure. Softening of the phonon spectrum should also lead to a reduction of the resistivity of cuprates under pressure, as observed in experiments [44, 45, 52, 53]. It is important, as our study has revealed, that this effect does practically not depend on the HTSC doping level. At the same time, the dependence $T_c(P)$ is strongly sensitive to the doping level suggesting another pressure effect on $T_c$. The increase of $T_c$ with pressure is most likely due to the mentioned above re-distribution of the charge carrier density $n_f$, resulting in the increase of $n_f$ in the conducting $CuO_2$ planes. As it has been mentioned above, this process is likely to take place easier in lightly doped cuprates [42, 44, 52, 53], where $dT_c/dP \approx +(4-5)\,\text{KGPa}^{-1}$ [26, 46, 53]. Thus, one may conclude that in optimally doped samples, pressure weakly affects $n_f$, which in optimally doped YBCO is close to saturation. Accordingly, in this case $dT_c/dP \approx +0.7\,\text{KGPa}^{-1}$, that is $T_c$ practically does not depend on pressure.

## 4. CONCLUSION

The effect of the hydrostatic pressure up to $0.95\,\text{GPa}$ on the excess conductivity $\sigma'(T)$ and the pseudogap $\Delta^*(T)$ of optimally doped $YBa_2Cu_3O_{6.94}$ single crystals with $T_c \approx 91.1\,\text{K}$ at $P = 0\,\text{GPa}$ has been studied. For the first time, it was observed that the ratio $D^* = 2\Delta^*(T_c)/k_B T_c$ as well as the pseudogap $\Delta^*$ increase at a rate $d\ln\Delta^*/dP = 0.32$, that is only a factor of 1.12 less than in lightly doped single crystals, implying almost similar influence of pressure on the coupling strength both in OD and SD YBCO. At the same time, the pressure effect on $T_c$, which is believed to be due to the rearrangement of the charge carrier density $n_f$ in the conducting $CuO_2$ planes, is minor, namely, $dT_c/dP \approx +0.73\,\text{KGPa}^{-1}$. One can therefore conclude that in optimally doped samples pressure feebly influences the charge carriers density $n_f$ in the $CuO_2$ planes, which in optimally doped YBCO is close to saturation. In consequence of this, $T_c$ is nearly independent of pressure. However, $d\ln\rho/dP \approx (-17 \pm 0.2)$ is found, which is of the same order of magnitude $\approx (-19 \pm 0.2)\,\text{GPa}^{-1}$ as in SD YBCO single crystals. Hence, the decrease of $\rho$ is not related to $n_f$, but is likely to occur in consequence of phonon spectrum softening in cuprates under pressure, that also explains the observed increase of both, $D^*$ and $\Delta^*(T)$ with increasing pressure. In this, a certain role can also be played by other specific mechanisms of the quasiparticle scattering [78–85] stipulated by the presence of the structural and the kinematic anisotropy in the system.

Independently of pressure, near $T_c$, $\sigma'(T)$ is described well within the framework of the 3D-AL and 2D-HL fluctuation theories with the MT contribution, demonstrating a 3D-2D crossover with increasing temperature. From the LP model analysis a rather peculiar temperature dependence of the pseudogap $\Delta^*(T)$ for the OD YBCO single crystal at pressures from $P = 0\,\text{GPa}$ to $P = 0.95\,\text{GPa}$ was found for the first time. The observed peculiarities allow one to comprehend better the way of the system behavior in the temperature range preceding the transition to the SC state. All these features can reasonably be accounted for in terms of the change of the local pair interaction with decreasing temperature. At the same time, independently of pressure, the resistivity curves only demonstrate a smooth decrease of $\rho(T)$ below $T^*$ without any noticeable features. This result suggests that the excess conductivity contains information about PG which can be deduced using our LP model approach. Thus, the analysis performed within the model, assuming the appearance of incoherent pairs (local pairs) in high-$T_c$ superconductors at $T \leq T^*$, has allowed us to get rather reasonable and self-consistent results for the temperature dependence of the pseudogap. At the same time, the pairing mechanism, which is responsible for the LP formation at such high temperatures, still remains controversial.